\newcommand{\CLASSINPUTbottomtextmargin}{25.4mm}
\newcommand{\CLASSINPUTtoptextmargin}{19.1mm}
\newcommand{\CLASSINPUToutersidemargin}{15.8mm}
\newcommand{\CLASSINPUTinnersidemargin}{17mm}
\documentclass[letterpaper,conference]{IEEEtran}
\ifCLASSINFOpdf
\else
\fi
\usepackage{amsmath}
\usepackage{siunitx}
\usepackage{graphicx}
\usepackage{calc}
\usepackage{algorithm}
\usepackage{algorithmic}
\usepackage{color}
\usepackage{array,multirow,makecell}
\usepackage{lineno}
\usepackage{color}
\usepackage{cite}
\usepackage[font=small]{caption}
\usepackage[font=small]{subcaption}
\usepackage{calc}
\makeatletter
\newcolumntype{P}[1]{>{\centering\arraybackslash}p{#1}}
\renewcommand\figurename{Figure}
\renewcommand{\appendixname}{Annex}
\setcellgapes{1pt}
\makegapedcells
\newenvironment{multicases}[1]
{\let\@ifnextchar\new@ifnextchar
\left\lbrace\def\arraystretch{1.2}%
\array{@{}l*{#1}{@{\quad}l}@{}}}
{\endarray\right.\kern-\nulldelimiterspace}
\makeatother
\newcommand{\RN}[1]{%
 \textup{\uppercase\expandafter{\romannumeral#1}}
}
\modulolinenumbers[5]
\begin{document}
\title{{\fontsize{24pt}{24pt}\selectfont
Scalable and Cost Efficient Algorithms for Virtual CDN Migration}\\}
\author{\IEEEauthorblockN{Hatem Ibn-Khedher\IEEEauthorrefmark{1},
Makhlouf Hadji\IEEEauthorrefmark{2},
Emad Abd-Elrahman\IEEEauthorrefmark{1}\IEEEauthorrefmark{4}, 
Hossam Afifi\IEEEauthorrefmark{1} and
Ahmed E. Kamal\IEEEauthorrefmark{3}}
\IEEEauthorblockA{\IEEEauthorrefmark{1}Institut Mines-Telecom (IMT), Telecom SudParis, Saclay, France. \IEEEauthorblockA{\IEEEauthorrefmark{4}Computer and Systems Department, National Telecommunication Institute (NTI), Cairo, Egypt.\\ Email: \{hatem.ibn\_khedher, emad.abd\_elrahman, hossam.afifi\}@telecom-sudparis.eu}}
\IEEEauthorblockA{\IEEEauthorrefmark{2}Technological Research Institute - IRT SystemX, 8 avenue de la vauve, 91120, Palaiseau, France.\\
Email: see http://makhlouf.hadji.free.fr/}
\IEEEauthorblockA{\IEEEauthorrefmark{3}Department of Electrical  Computer Engineering, Iowa State University, Ames, IA 50011-3060, USA.\\
Email: see http://www.ece.iastate.edu/~kamal/}
\thanks{This work is supported by the French FUI-18 DVD2C project: https://dvd2c.cms.orange-labs.fr/public-dvd2c/bienvenue-sur-le-site-du-projet-dvd2c.}}
\maketitle
\begin{abstract}
Virtual Content Delivery Network (vCDN) migration is necessary to optimize the use of resources and improve the performance of the overall SDN/NFV-based CDN function in terms of network operator cost reduction and high streaming quality. It requires intelligent and enticed joint SDN/NFV migration algorithms due to the evident huge amount of traffic to be delivered to end customers of the network. In this paper, two approaches for finding the optimal and near optimal path placement(s) and vCDN migration(s) are proposed (OPAC and HPAC). Moreover, several scenarios are considered to quantify the OPAC and HPAC behaviors and to compare their efficiency in terms of migration cost, migration time, vCDN replication number, and other cost factors. Then, they are implemented and evaluated under different network scales. Finally, the proposed algorithms are integrated in an SDN/NFV framework.
\end{abstract}

\begin{IEEEkeywords}
vCDN; SDN/NFV Optimization; Migration Algorithms; Scalability Algorithms.
\end{IEEEkeywords}
\IEEEpeerreviewmaketitle
\section{Introduction}
Software Defined Networks (SDN) and Network Function Virtualization (NFV) \cite{[1]} are two paradigms which aim to virtualize certain network functions while adding more flexibility and increasing the overall  network performance. ETSI has highlighted virtualized CDN (vCDN) among the major NFV use cases \cite{[2]}. The main purpose of vCDN is to allow the operator to dynamically deploy on demand virtual cache nodes to deal with the massive growing amount of video traffic. Scaling in/out, caching as a service etc.. are also among the key benefits of vCDN.

\begin{figure}[!b]
  \centering
\includegraphics[width=2.9in,height=1.9in]{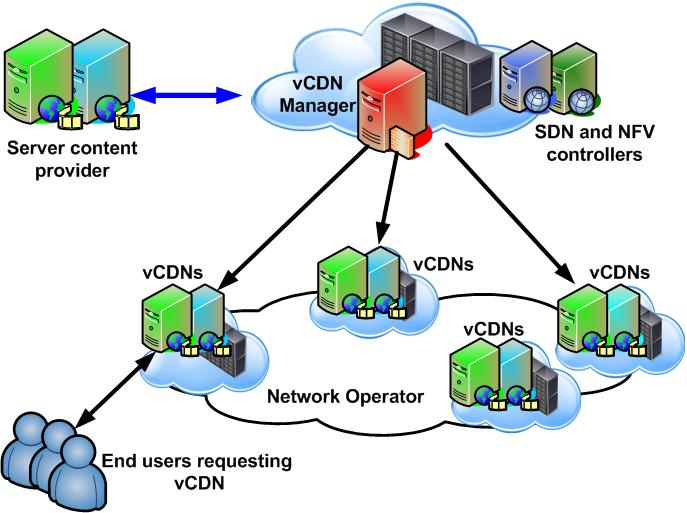}
 \caption {SDN/NFV-assisted CDN virtualization.}
 \label{vcdn}
\end{figure}

The overall virtualization challenges for CDN transition to vCDN are addressed in the research project DVD2C \cite{[3]}. In a previous work, we have investigated the main network issues in virtual machine migration and especially for vCDN use case \cite{[4]} and in an SDN/NFV optimization context \cite{[5]}. Fig. \ref{vcdn} depicts distributed vCDN nodes deployed by a centralized vCDN manager assisted with SDN/NFV controllers. Currently, despite the importance of optimization tasks, such functions are missing in the global architecture. We therefore, try in this paper to contribute with two algorithms and explain how they are integrated in the operators virtual LAN.
The proposed optimization algorithms in this paper consider vCDN migration problem inside the network operator. Moreover, the major objective from both algorithms is to minimize the total cost of content migration while minimizing the additional extra-costs needed for caching, streaming, and replication number. Through the first optimization (i.e. OPAC: Optimal Placement Algorithm for virtual CDN), we are going to formulate an exact algorithm based on a mathematical model for deciding the optimal location to migrate a vCDN or to instantiate (place) a new vCDN on demand to satisfy users quality requirements. Further, to cope with scalability problems of exact algorithms, we adapt a heuristic algorithm (i.e. HPAC: Heuristic Placement Algorithm for virtual CDN) to deal with our constraints when large scale networks need to be optimized. In this algorithm, we exploit the well known Gomory-Hu method to find a near to optimum point of operation. Finally, as the optimization algorithms deal with many heteroclite parameters, a clear view on how/where/when they are extracted and how they are reflected on a typical SDN/NFV architecture is diagrammed.

The rest of this paper is organized as follows: Section \RN{2}highlights the optimization algorithms in the virtualization context through the related work. Then, Section \RN{3}details OPAC parameters, constraints and its objective function. Section \RN{4}details our heuristic optimization algorithm (HPAC). Section \RN{5}evaluates the two algorithms and gives a comparison between them under predetermined metrics besides an integration diagram. Section \RN{6}concludes the work. 
\section{Related Work}
This section highlights the relevant NFV placement approaches either exact or heuristic as follows:  
Niels et al. \cite{[6]} provided a framework for multimedia delivery CDN-based on NFV. They used the three-layer structure of the network to seek the optimal locations of cache nodes. The optimization model lacks many  constraints such as  server's storage capacity.
    
Michele et al. \cite{[7]} used a mixed architecture where real and virtual CDN nodes coexist. Although they present an interesting hybrid solution, their architecture considers only the placement.
    
Hendrick et al. \cite{[8]} proposed a model for resource allocation of Virtual Network Functions (VNFs) within an NFV Infrastructure. Authors gave a hybrid solution inside the network operator that mixes the hardware and software-based network functions. Their model is a good abstraction of the deployment of NFV in an operator's network. But, it is still too general and does not address CDN specificity.
   
Bernardetta et al. \cite{[9]} raised an important network flow problem related to  compression/decompression processes applied on through-traffic passing by VNF instances. They propose a multi-level objective function for VNF placement optimization. Among the limitations of their approach is the utilization of a prioritization method to solve the problem which leads to sub-optimal results especially when the cost functions are orthogonal as in their case.

Mathieu et al. \cite{[10]} proposed a placement problem optimisation for the Deep Packet Inspection (DPI) networks through designing a virtualized DPI (vDPI) solution. Authors aim to propose a model capable of minimizing network load and  total number of deployed vDPI engines. This contribution is useful for our CDN placement problem. There are however major differences between the deployment of DPIs and CDNs. Moreover, authors do not introduce a migration scenario. 

Mathieu et al. \cite{[11]} tried in a second contribution, to solve the same aforementioned problem using heuristic algorithm which focuses on where to place the vDPI assisted with SDN paradigm. The proposed mechanism relies on a fitness function calculation to solve the problem.

Miloud et al. \cite{[12]} introduced the placement problem of virtual PDN/S-GW in the mobile core network. They propose three heuristic algorithms to solve the NFV placement problem based on applications and services types in the virtual instance selection process.

Further, in most of the above  exact/heuristic-based algorithms, the optimization of the virtual CDN migration is not  considered and, to our knowledge, no contribution  exploits the Gomory-Hu algorithm used in our solution as a scalable and a robust approach. 

\section{OPAC: Optimal Placement Algorithm for Virtual CDN}	
In this section, we specify the parameters and the constraints that are defined and proposed in formulating the optimization model OPAC. This formulation determines the migration of vCDN nodes to the optimal locations. We quote in Table \ref{Mathematical Notation} the main system and network parameters,  and decision variables. 

\begin{table}[!b]
\caption{Mathematical Notation}
\label{Mathematical Notation}
{\fontsize{8pt}{8pt}\selectfont
\renewcommand{\arraystretch}{1.3}
\begin{center}
\begin{tabular}{ l | p{5cm}}
\hline $\textbf{\textit{Parameters}}$& $\textbf{\textit{Definition}}$ \\
\hline ${V}$ & The set of client group nodes \\
\hline ${S}$ & The set of server nodes \\
\hline $D^{s}$ & Maximum throughput of the streaming server $s \in S$\\
\hline $F$ & The set of vCDN nodes\\
\hline $f_{size}$ & vCDN' s size $(vRAM, vCPU, vDISK)$ ($f \in F$)\\
\hline ${C}^{s}$ & Maximum storage capacity of the server $s$  \\
\hline $L_{i,j}$ &Link capacity between two nodes $i$ and $j$ (from $i$ to $j$)\\
\hline ${d}^f_{v}$ & Throughput used for streaming the functionality $f$ to the client group $v \in V$. It represents the user's demand requirements\\
\hline ${m}^{s}_{f}$ & The migration cost of the functionality $f$ to $s$ \\
\hline $\textbf{\textit{Decision \ variables}}$ & $\textbf{\textit{Definition}}$\\
\hline ${x}^{s}_{f}$ & Placement and migration binary variable which indicates that $f$ should move from origin server to optimal server $s$   \\
\hline ${y}^{s}_{v,f}$ & Binary variable which indicates the video hit from node $v$ of $f$ in server $s$ \\
\hline ${z}^{v,f}_{i,j}$ & Binary variable indicating whether the link $(i,j)$ is used to stream $f$ to $v$
 \\
\hline
\end{tabular}
\end{center}
}
\end{table}

\begin{itemize}
\item \textit{The decision variables}:
\begin{enumerate}
\item The binary variable ${x}^{s}_{f}$ indicates the placement of the streaming headend, and its migration from one server to another (best/optimal) location $s$. It is defined as:
\begin{equation}
{x}^{s}_{f} = 
   \begin{multicases}{2}
   1 & & \text{if $f$ migrates to $s$} \\
   0 & & \text{Otherwise}  \\
  \end{multicases}  \newline
\end{equation} 
\item The binary variable ${y}^{s}_{v,f}$ indicates a client $v$ needs a vCDN service, and the server $s$ caches it. It is defined as :
\begin{equation}
{y}^{s}_{v,f} = 
   \begin{multicases}{2}
   1      & & \text{if $v$ needs $f$ and $s$ caches $f$ } \\
   0 & & \text{Otherwise}  \\
   \end{multicases}  \newline
\end{equation}
\item The binary variable ${z}^{v,f}_{i,j}$ indicates whether a link $(i,j)$ is used (from $i$ to $j$) to stream from a server $s$ replicating $f$ (the one for which $y^{s}_{v,f} = 1$) to client $v$.  
\end{enumerate}
\item \textit{The constraints}:
\begin{enumerate}
\item The binary variable $y$ should be less than or equal to $x$. In fact, $y$ equals to 1, if and only if $v$ needs $f$, and $f$ is located on  server $s$. This means that $x$ should be equal to 1. Otherwise, if $y$ equals to 0, then $x$ may be equal to 0 or 1. 
\begin{equation}
\forall s \in S :{y}^{s}_{v,f} \leq {x}^{s}_{f} 
\end{equation} 
Note that a vCDN can be replicated more than once.
\item Only one optimal server should serve the client node that requests the functionality $f$:
\begin{equation}
\forall v \in V \mid {d}^f_{v} \neq 0 :
\sum_{s \in S} {y}^{s}_{v,f} = 1
\end{equation}
\item The cost of streaming $f$ by a server $s$ should be less than or equal to the maximum server capacity: 
\begin{equation}
\forall s \in S: \sum_{v \in V}\sum_{f \in F} {y}^{s}_{v,f} \times {d}^f_{v} \leq D^{s}
\end{equation} 
\item The storage of the optimal server should not exceed its maximum capacity:
\begin{equation}
\forall s \in S: \sum_{f \in F} {x}^{s}_{f} \times {f}_{size} \leq {C}^{s}
\end{equation}
\item Flow balance or conservation constraint between the server $s$ and the client node $v$ should be as the following: 
\begin{equation}
\sum_{j} {z}^{v,f}_{i,j} - \sum_{j} {z}^{v,f}_{j,i} =
\begin{multicases}{3}
   0 & \text{if } i \neq v , i \neq s\\
   {y}^{s}_{v,f} & \text{if } i = s\\
  -1 & \text{if } i = v \\
   \end{multicases}  \newline
\end{equation}
This is the network flow constraint. The sum of incoming flows must be equal to the outgoing ones.
\item Link capacity between source i and sink j should be larger than the flow on the link:
\begin{equation}
\forall i,j \in  V \cup S :
\sum_{v \in V}\sum_{f \in F} {z}^{v,f}_{i,j}\times {d}^f_{v} \leq {L}_{i,j}
\end{equation}
\end{enumerate}
The objective function is formulated in equation \eqref{eq:migration-cost} (cost function):
\begin{equation}
\min \sum_{s\in S}\sum_{f \in F} {x}^{s}_{f} \times {m}^{s}_{f}
\label{eq:migration-cost}
\end{equation} 

where ${m}^{s}_{f}$ is a parameter depending on the position of $s$, the position of $s_f$ (the server initially containing $f$ before migration), and the position of $s_v$ (the server initially connecting to the client group $v$). ${m}^{s}_{f}$ depends also on the size of $f$ and the operator policy. 
\end{itemize}

\begin{figure}[!t]
  \centering
 \includegraphics[width=2.5in]{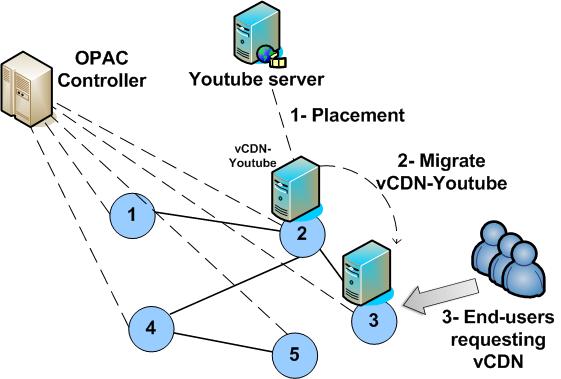}
 \caption{OPAC: vCDN-YouTube migration}
 \label{exampleopac}
\end{figure}

For the sake of clarity, we propose an example of OPAC vCDN migration as shown in Fig. \ref{exampleopac} . The algorithm is as follows:
1) As an input, the algorithm gathers SDN/NFV architectural information such as the initial placement of CDNs/vCDNs and all necessary dynamic parameters,
2) It needs also a prediction of end-users demands to execute OPAC, and
3) as a result, end-users requesting vCDN YouTube\textregistered , to watch a popular video in a live event for example, will be redirected to the videos hosted in the new optimal calculated location. 

Recall that the network operator hosting the vCDN of YouTube content provider migrates the vCDN cache/stream node to an optimal PoP where resources are available and content quality streamed are satisfied. 

The above problem is NP-hard due to our combinatorial complex system and therefore the proposed exact algorithm is difficult to scale up to decide where to deploy vCDN nodes in a large scale scenario. As a consequence, an efficient heuristic algorithm is proposed in the next section.
\section{HPAC: Heuristic Placement Algorithm for Virtual CDN}
\par Our proposal is strongly based on Gomory-Hu \footnote{noted by G-H in the rest of the paper} tree transformation \cite{[13]} of the initial network (represented by access, aggregate and core nodes) as shown in Fig. \ref{figure-network}). In other words, HPAC transforms the input network into a G-H tree. Then, replicating vCDNs is performed thanks to the G-H tree allowing to efficiently reduce the number of edges to be considered when migrating contents.

Algorithm \ref{HPAC-Algo} summarizes the pseudo code of HPAC. Hereafter, we describe these main stages.
\begin{algorithm}
{\fontsize{8pt}{8pt}\selectfont
\caption{\textbf{Heuristic algorithm for virtual CDN placement and migration }}
\begin{algorithmic}[1]
\STATE \textbf{Input:} ${V}$, ${S}$, $D^{s}$, $F$, $f_{size}$, ${C}^{s}$, $L_{i,j}$, ${d}^f_{v}$, ${m}^{s}_{f}$, $G=(V(G), E(G))$, \STATE ${s}_{v}$, ${s}_{f} $
\STATE \textbf{Output:} $x^s_{f}$, total migration cost
\STATE $GHT\leftarrow$ Gomory-Hu transformation \big($G$, $L_{i,j}$\big)
\STATE Tree-Exploration \big($GHT,{s}_{v},{s}_{f} $\big) 
\STATE Migration-process\big($D^{s},{C}^{s},L_{i,j},f_{size}, d^f_{v},{m}^{s}_{f}$\big):
\IF{$L_{i,j}, \leq d^f_{v}$}
\STATE Migrate-vCDN\big( \big)
\ENDIF
\end{algorithmic}
\label{HPAC-Algo}
}
\end{algorithm}
\subsection{Gomory-Hu Transformation}
At this stage, our heuristic HPAC computes the G-H tree of the CDN network or graph. The main advantage of G-H tree transformation is to compact this network graph structure using cuts to retain only feasible candidate topology and consequently lead to a smaller scale vCDN placement problem. We will insert a figure of the example of a G-H transformation.

For sake of clarity, we introduce the G-H tree transformation of our initial  graph $G=(V(G);E(G))$. The G-H tree $GHT=(V(GHT);E(GHT))$ of the former graph can be built using the definition in \cite{[14]}. There exists many algorithms that can build the G-H tree in polynomial times \cite{[15]} by finding $N-1$ maximum flow or minimum cuts between each pair of nodes in the graph, where $N=|V(G)|$. The G-H tree can be found in polynomial time according to the algorithm used to find a maximum flow/minimum cut in the initial graph. But, we describe only one selected strategy that relies on the minimum Steiner cut algorithm presented in \cite{[16]}. 

The transformation starts by initializing $V(GHT)$ to the set of the graph nodes (i.e., $\{V(G)\}$ and not $V(G)$). Concerning $E(GHT)$, it is initialized to an empty set. Besides, we define a queue list $Q$ to enable the G-H tree construction. $Q$ is initialized to $V(GHT)$ value. Then, while $Q$ is not empty, we pull the first element $S$ from this queue and we apply the minimum Steiner cut algorithm with the current set $S$ (i.e., first $Q$ element).

As the Steiner set in the new graph is obtained by contracting the entire sub-tree rooted at each neighbor in $GHT$ of $S$ into a single node. Consequently, two new sets of nodes $S_1$ and $S_2$ are generated from $S$ and $\lambda_{S_1,S_2}$ is the cut size. Accordingly, $V(GHT)$ should be updated by removing $S$ and adding $S_1$ and $S_2$. Besides, $E(GHT)$ should be enhanced by adding a new edge between $S_1$ and $S_2$ with capacity $\lambda_{S_1,S_2}$ (i.e., the cut size).

Furthermore, the queue $Q$ will be enriched by adding $S_1$ (respectively $S_2$) if it includes at least two nodes, $|S_1| > 1$ (respectively $|S_2| > 1$). The former steps will be repeated until the G-H tree construction matches with an empty queue $Q$ state.

The pseudo-code of the following G-H tree building strategy is summarized in Algorithm \ref{GH-Pseudocode}.
\begin{algorithm}
{\fontsize{8pt}{8pt}\selectfont
\caption{\textbf{Gomory-Hu tree transformation algorithm }}
\begin{algorithmic}[1]
\STATE \textbf{Input:} A connected graph $G=(V(G), E(G))$
\STATE \textbf{Output:} A tree $GHT=(V(GHT), E(GHT))$
\STATE $V(GHT)= V(G)$, 	$E(GHT)= \emptyset$, $Q=\{V(G) \}$
\WHILE{$Q\neq \emptyset$} 
\STATE $S\leftarrow$ Pull($Q$);//pull the first element from $Q$
\STATE $\{S_1,S_2\}\leftarrow$ minimum-Steiner-Cut($S, GHT$)
\STATE $V(GHT)= \{V(GHT)\setminus S\}\cup \{S_1, S_2\}$
\STATE $E(GHT)= E(GHT)\cup (S_1, S_2)$
\IF{$|S_1|> 1$}
\STATE5: $Q\leftarrow Q\cup S_1$
\ENDIF 
\IF{$|S_2|> 1$}
\STATE $Q\leftarrow Q\cup S_2$
\ENDIF
\ENDWHILE 
\end{algorithmic}
\label{GH-Pseudocode}
}
\end{algorithm}

\begin{figure}[t]
\begin{center}
\includegraphics[width=2.5in]{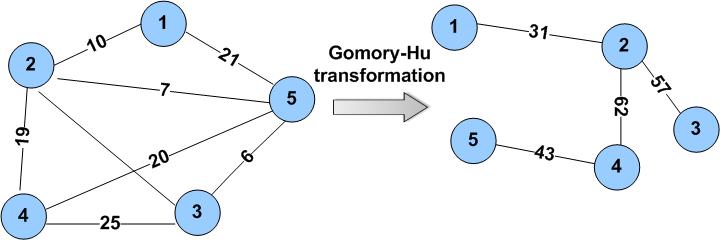}
\caption{Example of a Gomory-Hu tree transformation.}\label{figure-GH}
\end{center}
\end{figure}

We give a simple example of a G-H tree transformation of a graph $G$ as illustrated in Fig. \ref{figure-GH}. It is straightforward to see that the number of edges in the tree compared with the initial graph is reduced by $50\%$. Note that this G-H tree was built using only $6$ iterations of Algorithm \ref{GH-Pseudocode}. The Gomory-Hu approach consists to reduce considerably the complexity of the problem by reducing the number of edges to be considered in our optimization. In addition, it allows quantifying the amount of flows that can be transferred between each couple of nodes in the graph (see Fig. \ref{figure-GH}). As a consequence, this method facilitates and eliminates a large number of non-feasible solutions before the optimization.
\subsection{HPAC: Placement and Migration} 
Based on the G-H tree transformation, the number of nodes is equal to the number of nodes in the initial graph/network. However, the number of links in the derived G-H tree is at most equal to the number of nodes in the graph. Accordingly, the set of all paths  can be easily computed. In fact, in the tree structure there is only one path between any couple of servers.

With this tree transformation leading to significantly reduced input sizes, the problem of vCDN placement and migration will be easily solved even for large problem instances (large number of nodes and arcs). This is due to the efficiency of the G-H tree transformation leading to reduce considerably the domain of feasible solutions. So, it allows us to find the near optimal solutions in few seconds. In the following, we give more details on the second stage of the HPAC algorithm to place and migrate vCDNs in a cost efficient manner.

The second stage of HPAC algorithm consists to explore the unique path from an access point to the server containing the vCDN. Thus, if the  ingress flow cannot reach the destination (i.e. the vCDN), caused by a rupture node, then we will simply migrate the required vCDN to the rupture node of the G-H tree.

\begin{figure}[t]
\begin{center} \includegraphics[height=0.85in,width=2in]{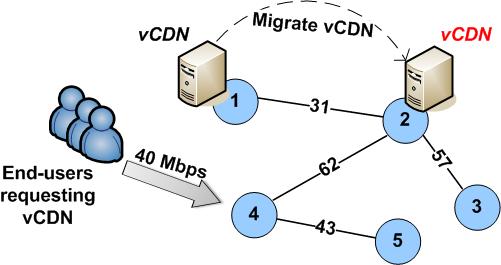}
\caption{Example of HPAC vCDN content replication/migration.}
\label{example-migration}
 \end{center}
\end{figure}

We propose the following example (see Fig. \ref{example-migration} ) to illustrate HPAC algorithm for migrating and placing judiciously vCDNs. It depicts a scenario of content replication/migration. In this scenario, a client cluster (group) representing grouped end-user requests for a vCDN with a specific throughput (i.e., content quality) equal to $40 \ Mbps$. The proposed migration method searches to migrate the desired vCDN from its initial deployed position (server 1) to the suitable server (i.e., server 2 which has more than the required streaming capacity) in an efficient way (i.e., with minimum migration cost). 

\section{OPAC versus HPAC (Exact versus Heuristic)}
For the interest of assessing the efficiency of OPAC and HPAC algorithms, we used \cite{[17]} and \cite{[18]} as optimization tools. In addition, different metrics/cost-factors can be defined as follows:
\begin{equation}
Migration \ cost = \sum_{s\in S}\sum_{f \in F} {x}^{s}_{f} \times {m}^{s}_{f}
\end{equation}
\begin{equation}
Migration \ time = \sum_{s\in S}\sum_{f \in F} {x}^{s}_{f}\times{f}_{size}\times\max_{(i,j)\in P_{s_{f}, s}}\frac{1}{L_{i,j}}
\label{eq:migration}
\end{equation}

Where $P_{s_{f}, s}$ is a given path from $s_{f}$ to $s$. In fact, we are here assuming that the migration is done in a sequential way. If the migration is performed in parallel, then the migration time would be given by: $
 \max\limits_{s\in S, f \in F} \left({x}^{s}_{f}\times{f}_{size}\times\max_{(i,j)\in P_{s_{f}, s}}\frac{1}{L_{i,j}}\right)
$. However, we will only consider \eqref{eq:migration}.
\begin{equation}
\hspace{-0.25in}
Replica \ number = \sum_{s \in S\backslash\lbrace{s_{f}}\rbrace}\sum_{f \in F} x^{s}_{f}
\end{equation}
\begin{equation}
\centering
vCache \ cost =\frac {\sum_{s \in S} \sum_{f \in F} {x}^{s}_{f} \times {f}_{size}} {\sum_{s \in S} C^{s}}
\end{equation}
\begin{equation}
vStream \ cost =\frac {\sum_{v \in V}\sum_{f \in F} {y}^{s}_{v,f} \times {d}^f_{v}} {\sum_{s \in S} D^{s}}
\end{equation}

Where the cost represents the amount of system resources (vCache) or network resources (vStream) consumed when migrating a vCDN from a server to another one.

Moreover, in order to decide which algorithm should we use and when, different network scales (i.e. small and large) and topologies are considered as follows:
\subsection{Small Scale Scenario: a network operator snapshot}
\begin{figure}[!t]
\centering
\includegraphics[height=2in,width=2.5in]{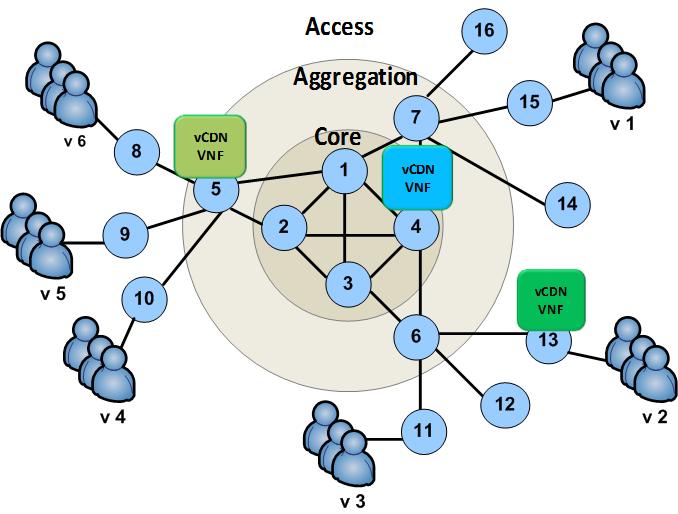}
\caption{Network topology used for small scale.}
\label{figure-network}
\end{figure}

In this subsection, the optimization targets a small number of vCDNs. Therefore, a snapshot of a three-tier network operator architecture \footnote{NFVI PoP Network Topology (2016): https://tools.ietf.org/pdf/draft-bagnulo-nfvrg-topology-01.pdf}  is used for the evaluation as show in Fig. \ref{figure-network}. Moreover, the main dissimilarities between the two approaches according to the defined metrics are showed hereafter:

\textbf{Algorithm complexity:} The algorithmic complexity of G-H-based HPAC is polynomial while it is exponential in OPAC. Table \ref{Efficiency comparison between OPAC and HPAC} illustrates and defines the computing complexity of the proposed algorithms OPAC/HPAC. It is clear that the exact formulation of the problem has an exponential number of feasible solutions defined by the convex hull of the discussed problem. To cope with scalability issues, we investigated a heuristic solution based on Gomory-Hu approach to reduce considerably the complexity problem as shown in Table \ref{Efficiency comparison between OPAC and HPAC}.

\begin{figure}[!t]
\centering
\begin{subfigure}[!t]{0.25\textwidth}
\includegraphics[height=1.5in,width=2.8in]{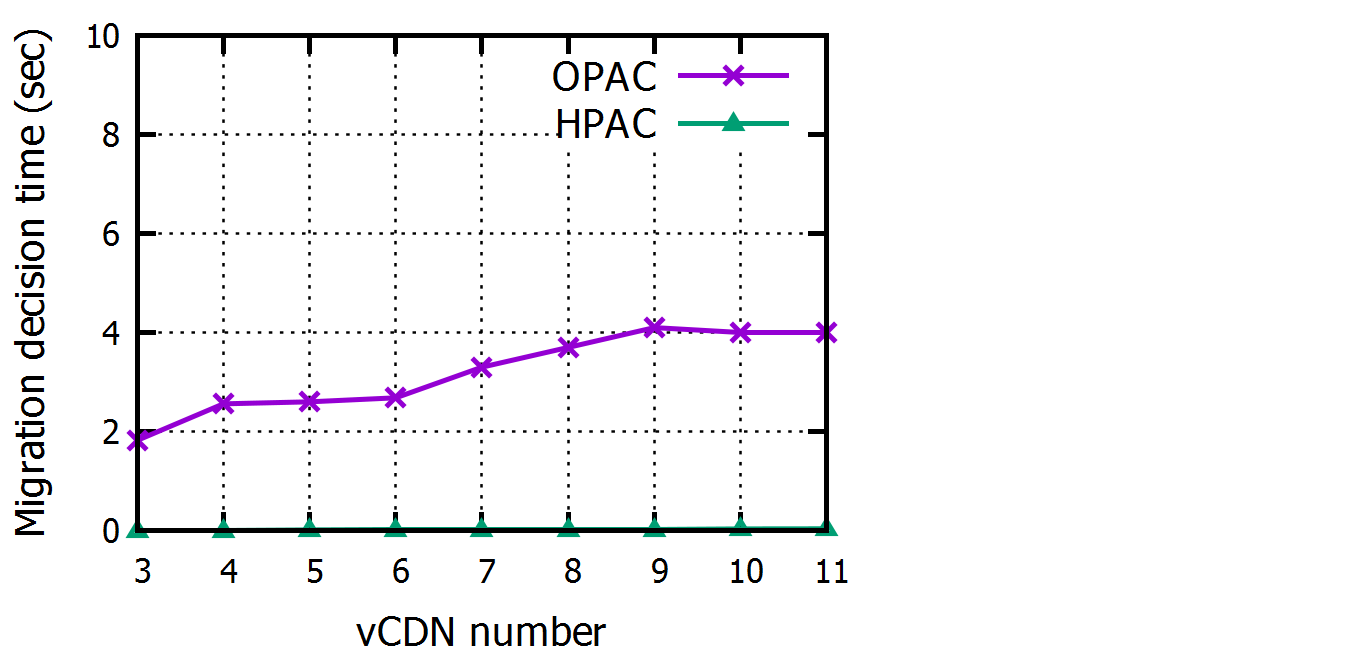}
\caption{Migration decision time.}
\label{mig:decision:time}
\end{subfigure}
\begin{subfigure}[]{0.25\textwidth}
\includegraphics[height=1.5in,width=2.8in]{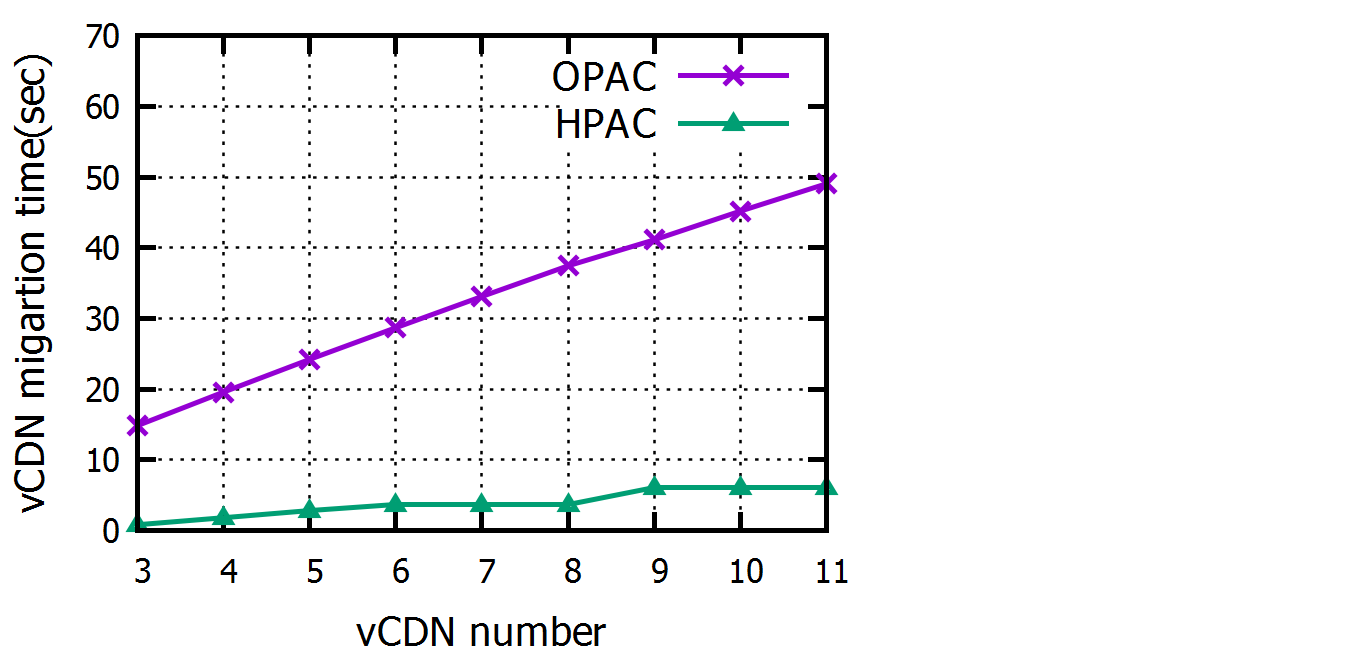}
\caption{Migration time.} 
\label{mig:time}
\end{subfigure}
\begin{subfigure}[]{0.25\textwidth}
\includegraphics[height=1.5in,width=2.8in]{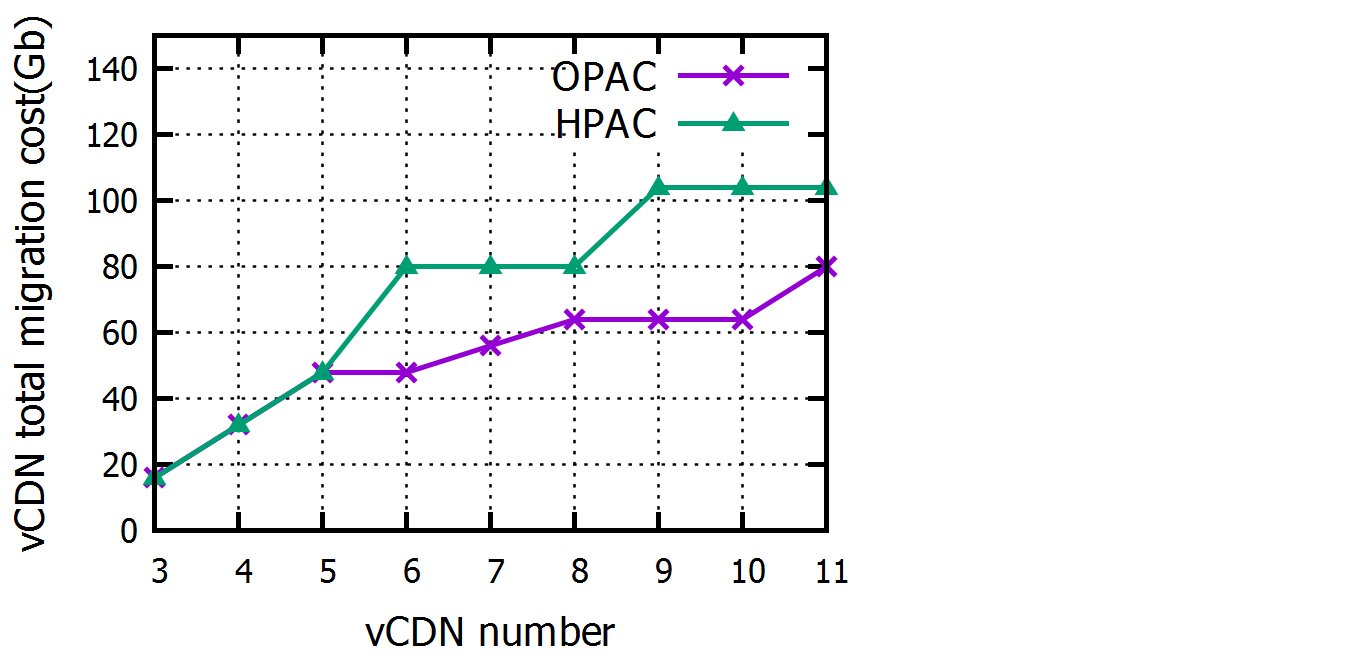}
\caption{vCDN total migration cost.}
\label{mig:migration:cost}
\end{subfigure}
\begin{subfigure}[]{0.25\textwidth}
\includegraphics[height=1.5in,width=2.8in]{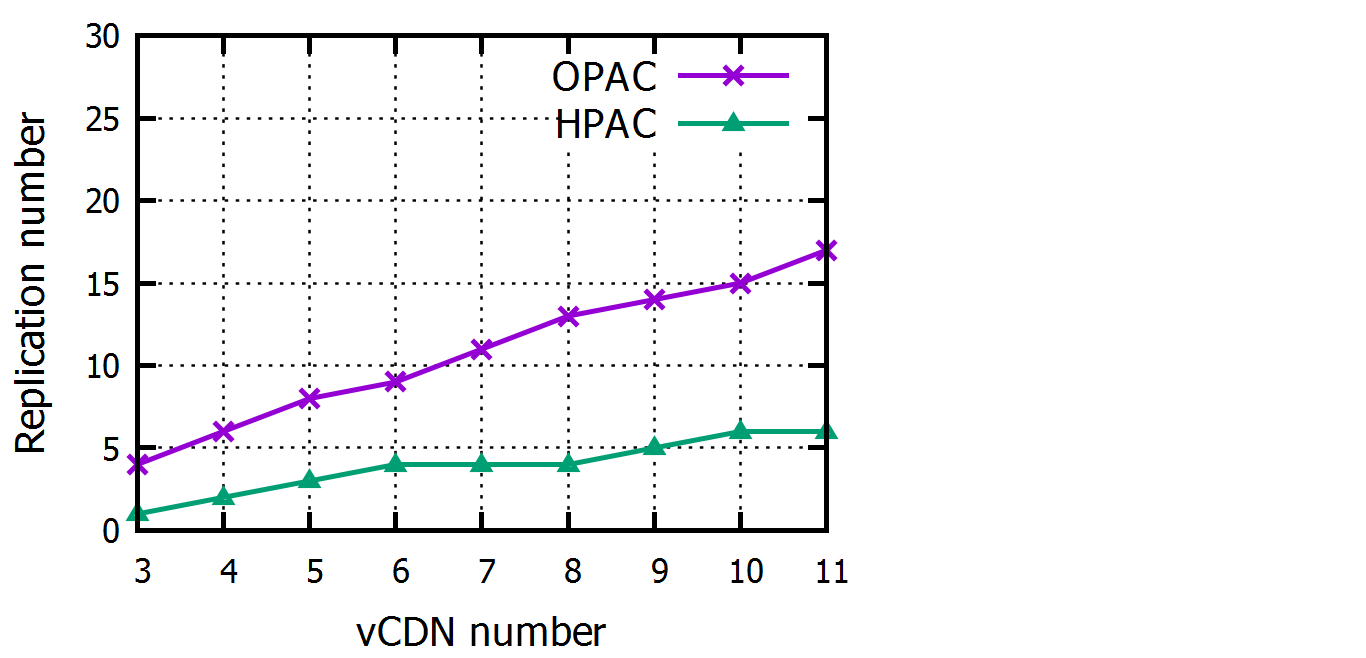}
\caption{Replication number.}
\label{replication_number}
\end{subfigure}
\begin{subfigure}[]{0.25\textwidth}
\includegraphics[height=1.5in,width=2.8in]{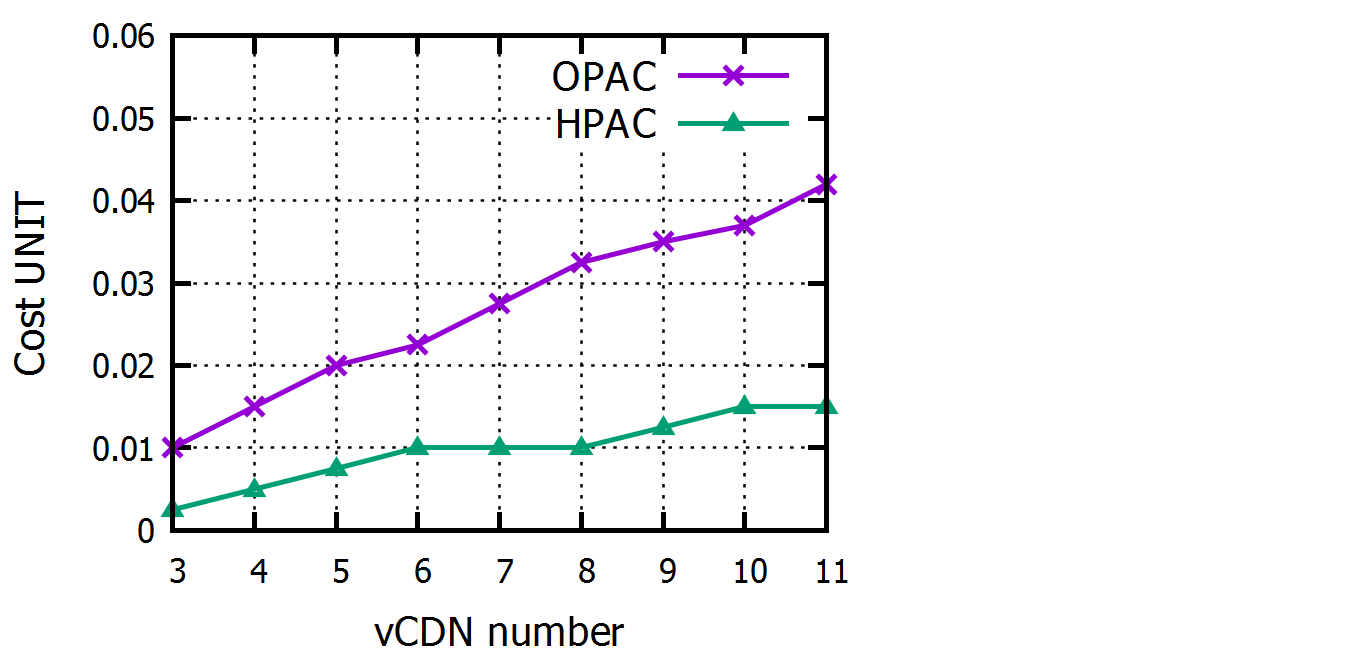}
\caption{Virtual cache.}
\label{vcache}
\end{subfigure}
\begin{subfigure}[]{0.25\textwidth}
\includegraphics[height=1.5in,width=2.8in]{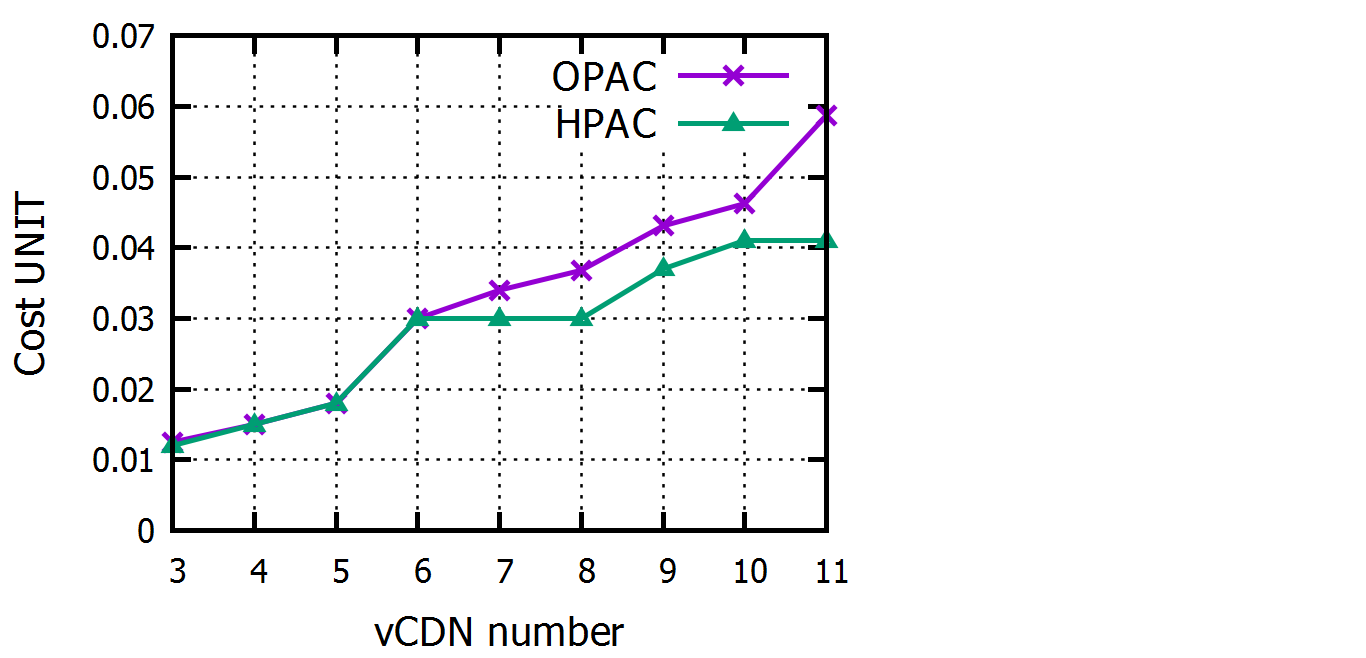}
\caption{Virtual stream.}
\label{vstream}
\end{subfigure}
\caption{OPAC-HPAC comparison in the small network scale scenario.}
\end{figure}

\textbf{The run-time:} The time convergence of the two algorithms is shown in the Fig. \ref{mig:decision:time} . HPAC outperforms OPAC since the latter is a combinatorial-based algorithm and has an exponential complexity.

\textbf{The total migration time:} It is the duration needed to migrate a vCDN to the optimal/near optimal point of the deployment. Fig. \ref{mig:time} shows the average migration time needed for migration vCDN delivery functions to the optimal instantiation point taking into account the NFV constraints including system, network, and content quality parameters which are related to the vCDN functionalities. Although HPAC gives the shortest time as depicted in this figure, OPAC is still acceptable. Indeed, This metric is strongly related to the distance between servers as written in \eqref{eq:migration} and the time consumed for the vCDN migration process is less than $1 minute$ for vCDN numbers equals $11$.

\textbf{The total migration cost:} It is evaluated in both approaches (exact and heuristic). In Fig. \ref{mig:migration:cost} , the vCDN total migration cost is measured in terms of $ Gigabits$ $(Gb)$ against vCDN number. The Fig. \ref{mig:migration:cost} shows that vCDN-migration cost is increasing in both approaches for vCDN number ranging from 3 to 6. This is due to the non uniform client group distribution. It is noticeable that OPAC outperforms HPAC for vCDN ranging from 6 to 11. Nevertheless, HPAC proofs a low variation cost which leads to a better save of operator resources. 

\textbf{The total replication number:} Under a random end-users matrix demand, the total replication number in both approaches is not significant (small) as shown in Fig. \ref{replication_number}. However, HPAC is the strictest approach when replicating vCDN in small scale scenario. 

\textbf{Other cost-factors:} OPAC and HPAC are compared in terms of virtual cache (vCache) and virtual stream (vStream) cost-units \footnote{Cost-unit may be defined by the utilization percentage of the available system or network resource. It is the amount of resources (in GBytes) consumed when migrating content from a server to another one.}. In Fig. \ref{vcache} cost unit is plotted against vCDN number. The virtual in-networking caching cost increases in both algorithms but they are still insignificant.  Similarly, in Fig. \ref{vstream} the virtual in-network streaming cost increases with vCDN number in both algorithms but it is still insignificant. Although the HPAC is slightly cost-efficient, both algorithms proofed an efficient network resource saving.

Experiments show that OPAC is still acceptable and saves about 96 \% of the system resources and 94\% of the network resources at $\left\vert{F}\right\vert = 11$. Nevertheless, it is still the most costly comparing to HPAC in terms of system and network resource usage.

For the sake of clarifying the effectiveness of the HPAC, we define the Gap metric as follows:
\begin{equation}
Gap (C) = \frac{C_{HPAC} - C_{OPAC}}{ C_{OPAC}}
\label{eq:gap}
\end{equation} 

Where $C$ is a cost factor (e.g., migration cost).

Table \ref{Gap} depicts the Gap metric (\%) calculation formulated in \eqref{eq:gap} to estimate the difference in cost between OPAC and HPAC. It demonstrates the efficiency of the HPAC algorithm in terms of total migration cost since the Gap values are close to zero. In other words, the objective function of our heuristic solution is always very close to the optimum. This justifies the efficiency of HPAC. 

\begin{table}[!t]
{\fontsize{8pt}{8pt}\selectfont
\renewcommand{\arraystretch}{1.3}
\caption{Gap (migration \ cost): HPAC efficiency}
\label{Gap}
\begin{center}
\begin{tabular}{p{1.7cm}|p{0.5cm}|p{0.5cm}|p{0.5cm}|p{0.5cm}|p{0.5cm}|p{0.5cm}}
\hline $\textbf{\textit{vCDN number}}$&$6$&$7$&$8$&$9$&$10$&$11$\\
\hline $\textbf{\textit{Gap (\%)}}$&$0.66$&$0.42$&$0.25$&$0.62$&$0.62$&$0.3$\\
\hline 
\end{tabular}
\end{center}
}
\end{table}
\subsection{Large Scale Scenario: an Erdos-Renyi graph-based network operator}
\begin{figure}[!t]
 \centering
 \begin{subfigure}[!t]{0.25\textwidth}
 \includegraphics[width=1.5in]{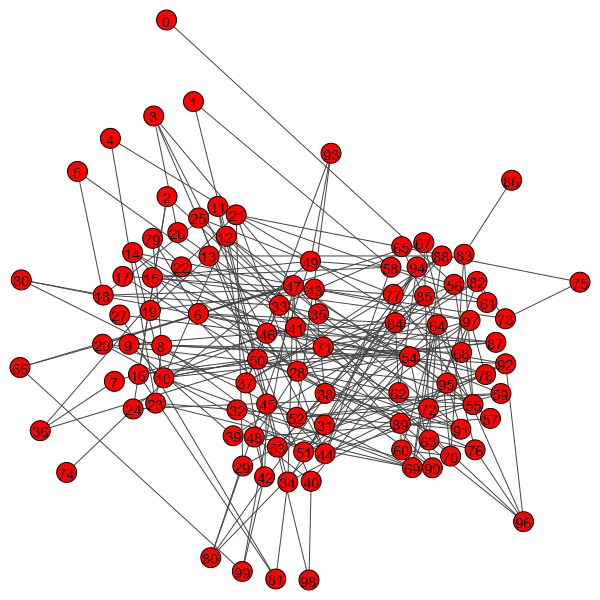}
 \caption{Network topology.}
 \label{erg}
 \end{subfigure}%

\begin{subfigure}[]{0.25\textwidth}
\includegraphics[width=1.5in]{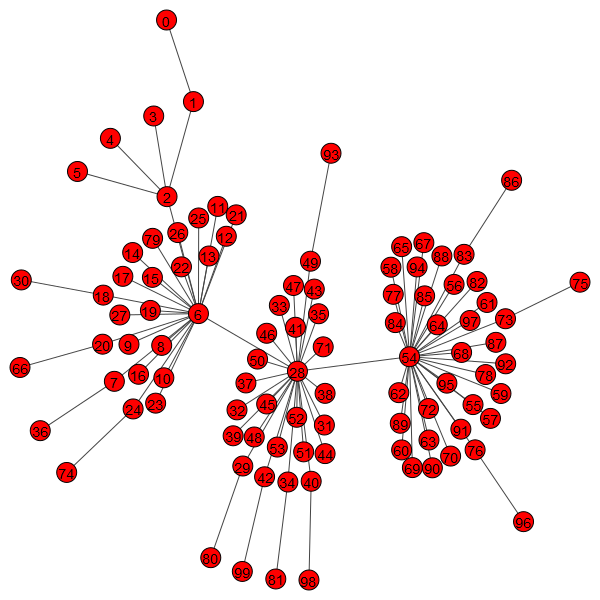}
\caption{G-H-based transformation.}
\label{gh_erg}
\end{subfigure}
\caption{Network topology used for large scale.}
\end{figure}

In this subsection, the optimization targets a large number of vCDNs (\si\micro\ vCDNs/containers). The scenario relies on the well known Erdos-Renyi \footnote{Erdos-Renyi graphs are random topologies that can be used to assess performance of our algorithms using random connectivity: one of real life scenarios. This topology is very similar to those generated by GT-ITM tool. Our objective in using this topology is to address well known random topologies to better test our solutions.} undirected and weighted graph. The graph has 100 vertices (nodes) and 200 edges as shown in Fig. \ref{erg}. Its G-H-based transformation is shown in Fig. \ref{gh_erg} which has only 99 edges ($49.5 \%$). These figures depict the topology used for evaluating HPAC algorithm in a large scale scenario. The OPAC algorithm could not be used here with reasonable resources.

\begin{figure}[!t]
\centering
\begin{subfigure}[]{0.25\textwidth}
\centering
\includegraphics[height=1.5in,width=2.7in]{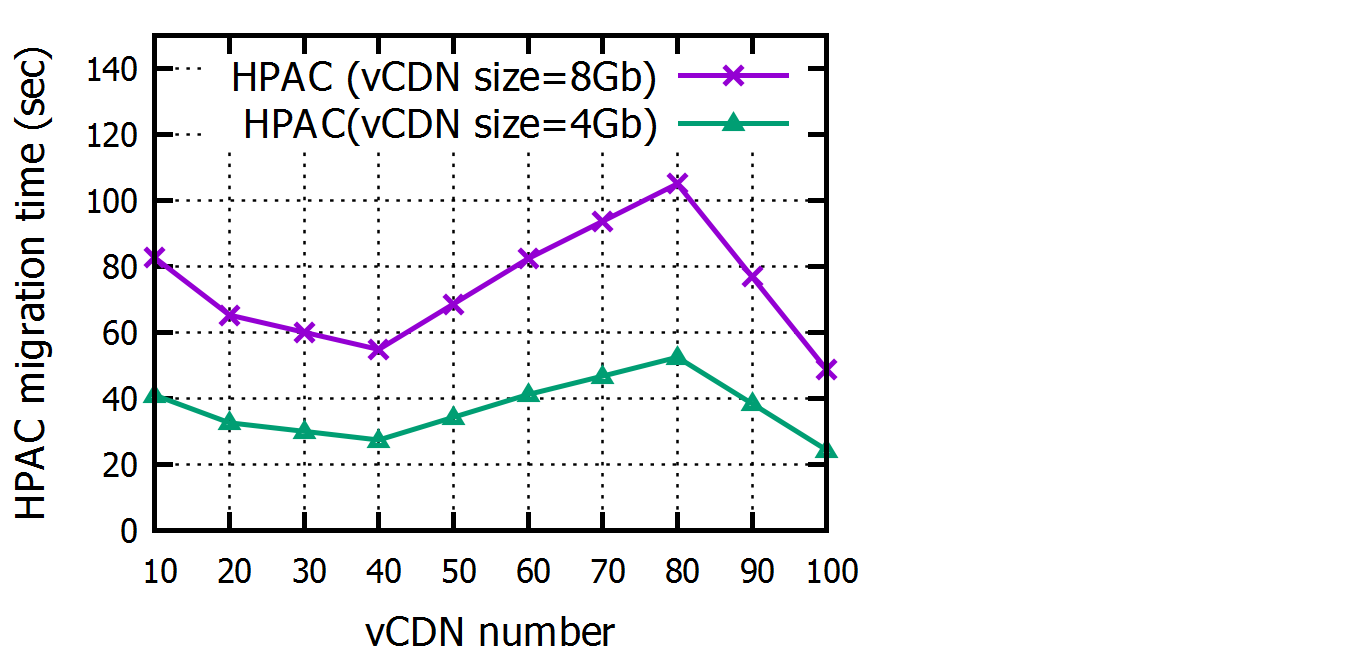}
\caption{Migration time/delay.}
\label{hpacmd}
\end{subfigure}
\begin{subfigure}[]{0.25\textwidth}
\centering
\includegraphics[height=1.5in,width=2.7in]{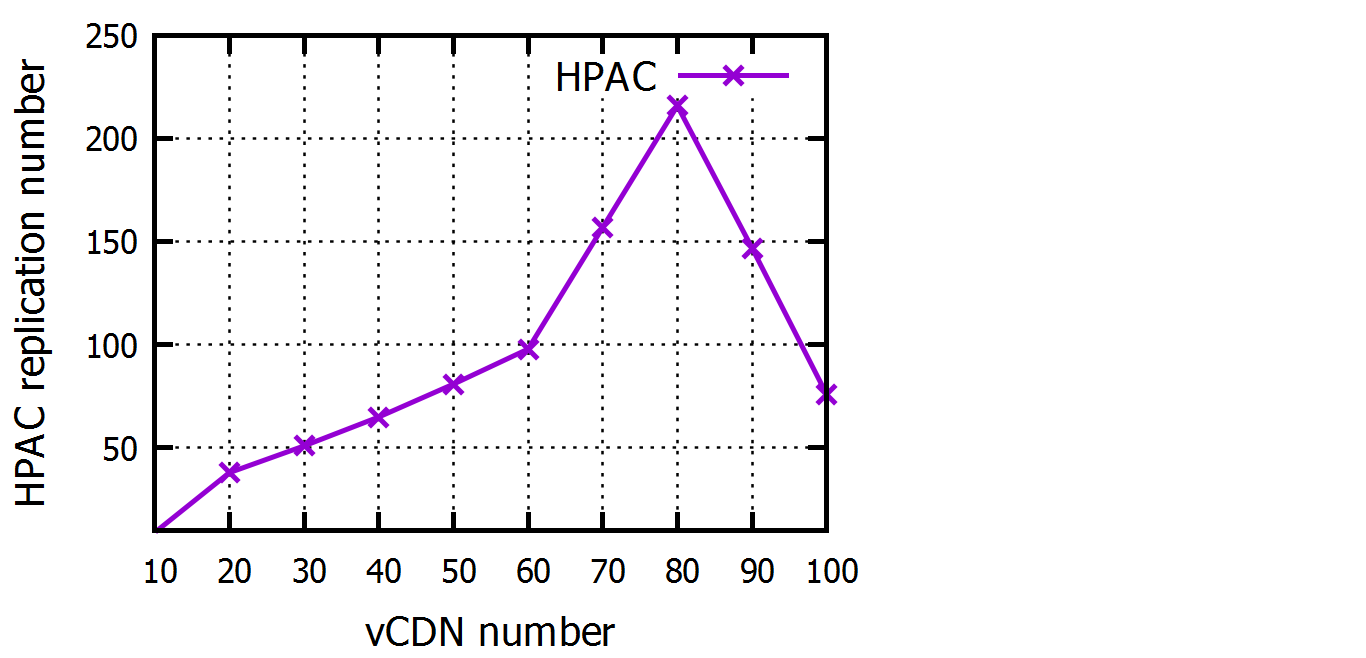}
\caption{Replication number.}
\label{hpacrn}
\end{subfigure}
 \begin{subfigure}[]{0.25\textwidth}
 \centering
\includegraphics[height=1.6in,width=2.8in]{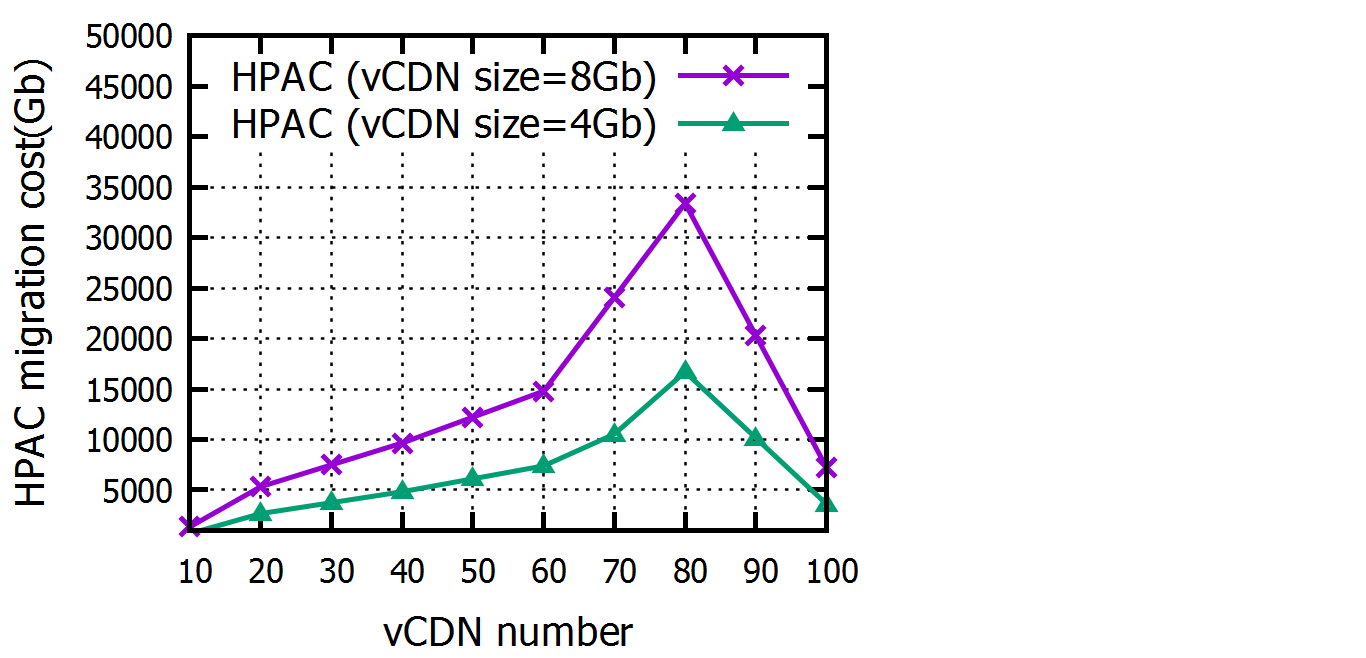}
\caption{Migration cost.}
\label{hpacmc}
\end{subfigure}
\caption{HPAC in large network scale.}
\end{figure}

\begin{figure}[!t]
\centering
\hspace*{0.3in}\includegraphics[height=2in,width=4in]{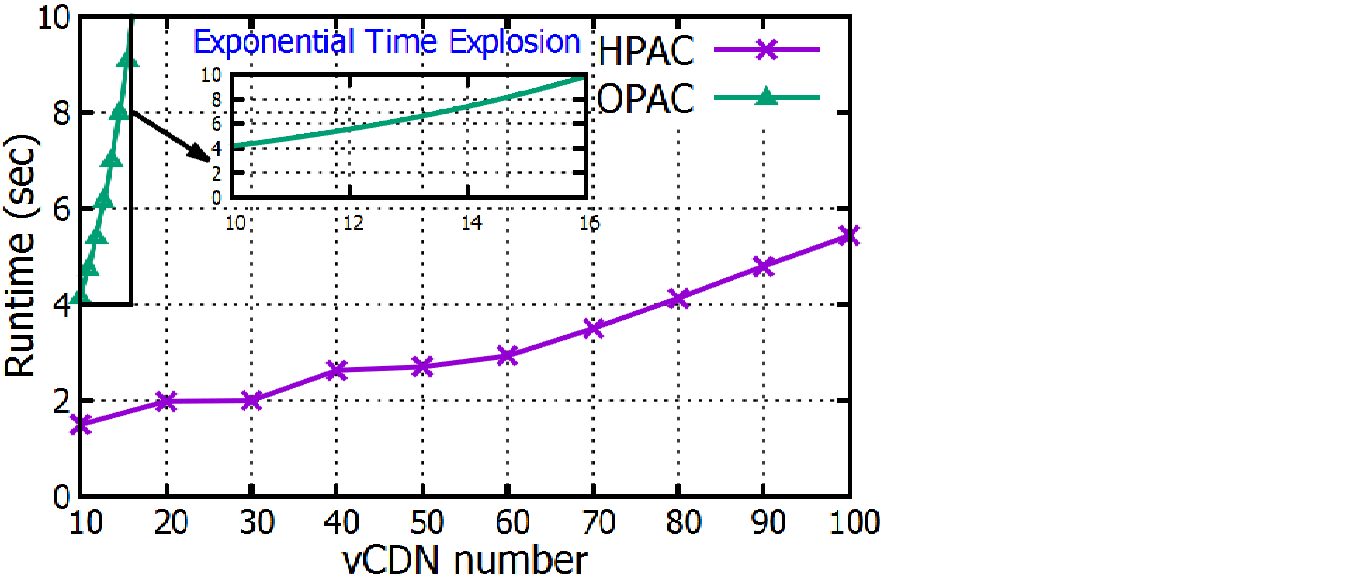}
\caption{OPAC and HPAC run-time in large network scale.}
\label{hpacrt}
\end{figure}

Fig. \ref{hpacmd} and \ref{hpacmc} depict the total migration delay time and the total migration cost respectively. Further, Fig. \ref{hpacrn} shows the total replication number. These metrics are increasing until a decreasing slope at $\left\vert{F}\right\vert=80$ representing the average vCDN number stabilizing the system. Furthermore, Fig. \ref{hpacrt} shows the run-time of the OPAC and HPAC approaches in a large scale scenario. It demonstrates the feasibility/efficiency of HPAC since its run-time is in terms of a few seconds $(6 \ sec)$ while OPAC run-time explodes from vCDN number equals 20.

\subsection{Interpretations}
In this subsection, we quote the interpretations that can be revealed from the results above:

\textbf{In small scale scenario:} (the order of ten servers, ten client groups, and ten vCDNs), the exact approach is suggested to be used because its run-time is in terms of a few seconds (see Fig. \ref{mig:decision:time}. OPAC is what we should choose as an optimization mechanism because it gives the lowest migration cost as shown in Fig. \ref{mig:migration:cost} .

\textbf{In large scale scenario:} (the order of hundred servers, hundred client groups, and hundred vCDNs), the exact approach is suggested to be useful and the G-H-based HPAC algorithm is the alternative to solve the vCDN migration problem since its algorithm run-time is in terms of a few seconds as shown in Fig. \ref{hpacrt} .

A brief comparison between the two approaches which answers implicitly to the question when to use OPAC and when to use HPAC is given in the Table \ref{Efficiency comparison between OPAC and HPAC} .

\begin{table}[!b]
{\fontsize{8pt}{8pt}\selectfont
\renewcommand{\arraystretch}{1.3}
\caption{Efficiency comparison between OPAC and HPAC; SS: Small Scale, LS: Large Scale $N=\left\vert{V(G)}\right\vert$, $M=\left\vert{E(G)}\right\vert$}
\label{Efficiency comparison between OPAC and HPAC}
\begin{center}
\begin{tabular}{P{1.7cm}|P{2.6cm}|P{2cm}}
\hline$\textbf{\textit{Metrics}}$&$\textbf{\textit{OPAC}}$&$\textit{\textbf{HPAC}}$ \\
\hline \textit{Algorithm complexity}&Exponential &$\mathcal{O}(N^{3}*M^{1/2})$\\
\hline \textit{Run-time}& A few seconds in SS; a few minutes in \textit{LS}&A few second in \textit{SS/LS}\\
\hline \textit{Total migration cost} &Excellent in SS;unfeasible in \textit{LS}& Good in \textit{SS/LS}\\
\hline\textit{ Total migration time (i.e.,delay)}&Good in \textit{SS};unfeasible in \textit{LS}&Excellent in both \textit{SS/LS}\\
\hline \textit{Replication number}&Free/loose in \textit{SS}; unfeasible in \textit{LS}& Free/loose in \textit{SS/LS}\\
\hline \textit{vCache/vStream cost} &Good minimization in \textit{SS}& Excellent minimization in both \textit{SS/LS}\\
\hline 
\end{tabular}
\end{center}
}
\end{table}

For the sake of clarifying better the efficiency of our solutions, a brief comparison between our algorithms (OPAC/HPAC) and the state-of-the-art is shown in Table \ref{Comparison between OPAC/HPAC and state-of-the-art}. We proposed an exact method based on linear integer model that will always find the optimal solution (for small and medium instances of the problem). This exact model is used as a benchmark (the best point of comparison) of a novel heuristic algorithm, elaborated to cope with scalability issues of the problem. Nevertheless, we also compared in Table \ref{Comparison between OPAC/HPAC and state-of-the-art} our approaches to similar solutions in the literature. 

\begin{table}[!b]
\centering
{\fontsize{8pt}{8pt}\selectfont
\begin{center}
\caption{Comparison between OPAC/HPAC and state-of-the-art}
\begin{tabular}{ P{0.6in} |P{0.5in}|p{0.6in}|p{0.8in}}
\hline$\textbf{\textit{Work}}$&$\textbf{\textit{Approach}}$&$\textit{\textbf{Metrics}}$&$\textit{\textbf{Limitations}}$ \\
\hline \vspace{0.17in}  Niels et al. \cite{[6]}&\vspace{0.23in}Exact&\vspace{0.05in}Bandwidth, deployment cost & No virtual node cooperation, no QoE parameters, missed capacity and RAM constraints \\
\hline \vspace{0.05in} Michele et al. \cite{[7]}&\vspace{0.1in}Exact&\vspace{0.1in}Usage cost& Known traffic distribution (static), unreliable\\
\hline \vspace{0.17in} Hendrick et al. \cite{[8]} &\vspace{0.22in}Exact&\vspace{0.05in}Deployment cost, service request &  Virtual CPU, virtual storage, virtual capacity, and QoE constraints are missed \\
\hline \vspace{0.19in} Bernardetta et al. \cite{[9]}&\vspace{0.3in}Exact&\vspace{0.04in}Allocated computing resources, link utilization&  Prioritization method with orthogonal cost functions, virtual RAM and QoE constraints are missed \\
\hline \vspace{0.05in} Mathieu et al. \cite{[10]}&\vspace{0.1in}Heuristic&\vspace{0.02in}CAPEX, server's load &\vspace{0.06in} Missed NFV constraints\\
\hline \vspace{0.05in} Mathieu et al. \cite{[11]}&\vspace{0.08in}Exact and Heuristic&\vspace{0.01in}CAPEX, server's load & \vspace{0.01in}Offline measurement, and no migration \\
\hline \vspace{0.05in} Miloud et al \cite{[12]}&\vspace{0.1in}Exact and Heuristic&\vspace{0.1in}Traffic's load &Lacks system constraints and there is no VNF (PDN-GW) migration\\
\hline \vspace{0.07in} OPAC\&HPAC algorithms&\vspace{0.05in}Exact and Heuristic& System, network, and quality metrics&\vspace{0.10in}No limitations\\
\hline 
\end{tabular}
\label{Comparison between OPAC/HPAC and state-of-the-art}
\end{center}
}
\end{table}

\subsection{Integration of the Algorithms}
To our knowledge, the proposed optimization algorithms have to be integrated in a vCDN controller (as seen Fig. \ref{exampleopac} ). It interacts with the vCDN manager software in a legacy NFV-MANO framework \cite{[19]}. Further, as the migration problem deals with different heteroclite parameters and variables, a clear view about how they are extracted and reflected on an SDN/NFV framework is necessary.

\begin{figure}[!t]
\begin{center}
\includegraphics[height=2in,width=3in]{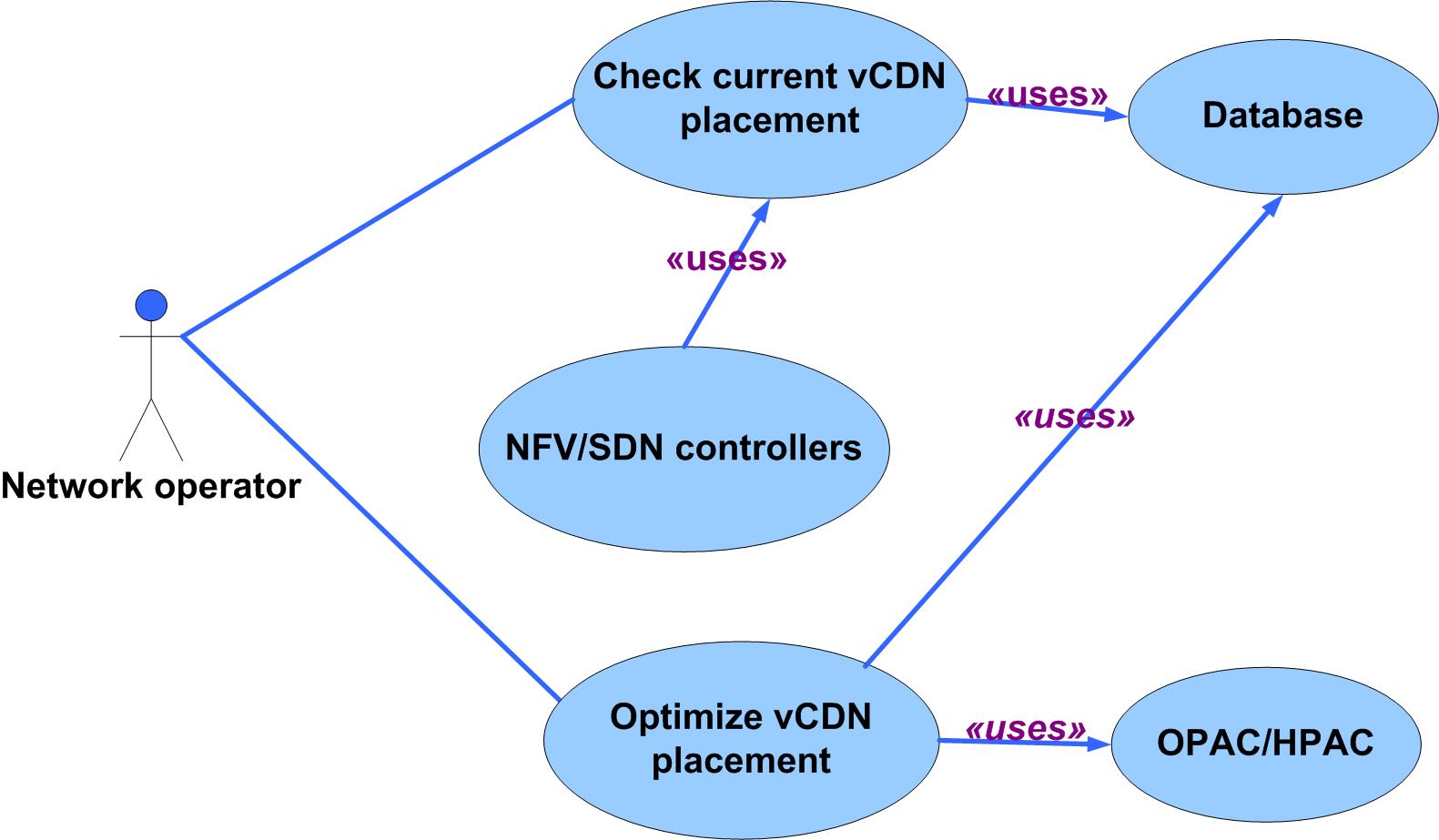}
\caption{Main use-cases for OPAC/HPAC in an SDN/NFV framework.}
\label{use-cases}
\end{center}
\end{figure}

In Fig. \ref{use-cases} and for the sake of simplicity, the two main use cases needed for integrating OPAC/HPAC algorithms by the network operator are depicted. Indeed:
\begin{itemize}
\item Use case 1: The network operator checks the current placement of vCDNs. To do this, he queries a Database using SQL structured language and an NFV/SDN controllers (e.g., Openstack \cite{[20]}/Opendaylight \cite{[21]}) using API interfaces. Openstack horizon and Opendaylight DLUX provide respectively the system and network resource information needed to complete the operator database.
\item Use case 2: The network operator requests for the vCDN optimization placement/migration result using either OPAC or HPAC according to the actual network scale (small or large).
\end{itemize}

\begin{figure}[!t]
\begin{center}
 \includegraphics[height=2.5in,width=3.5in]{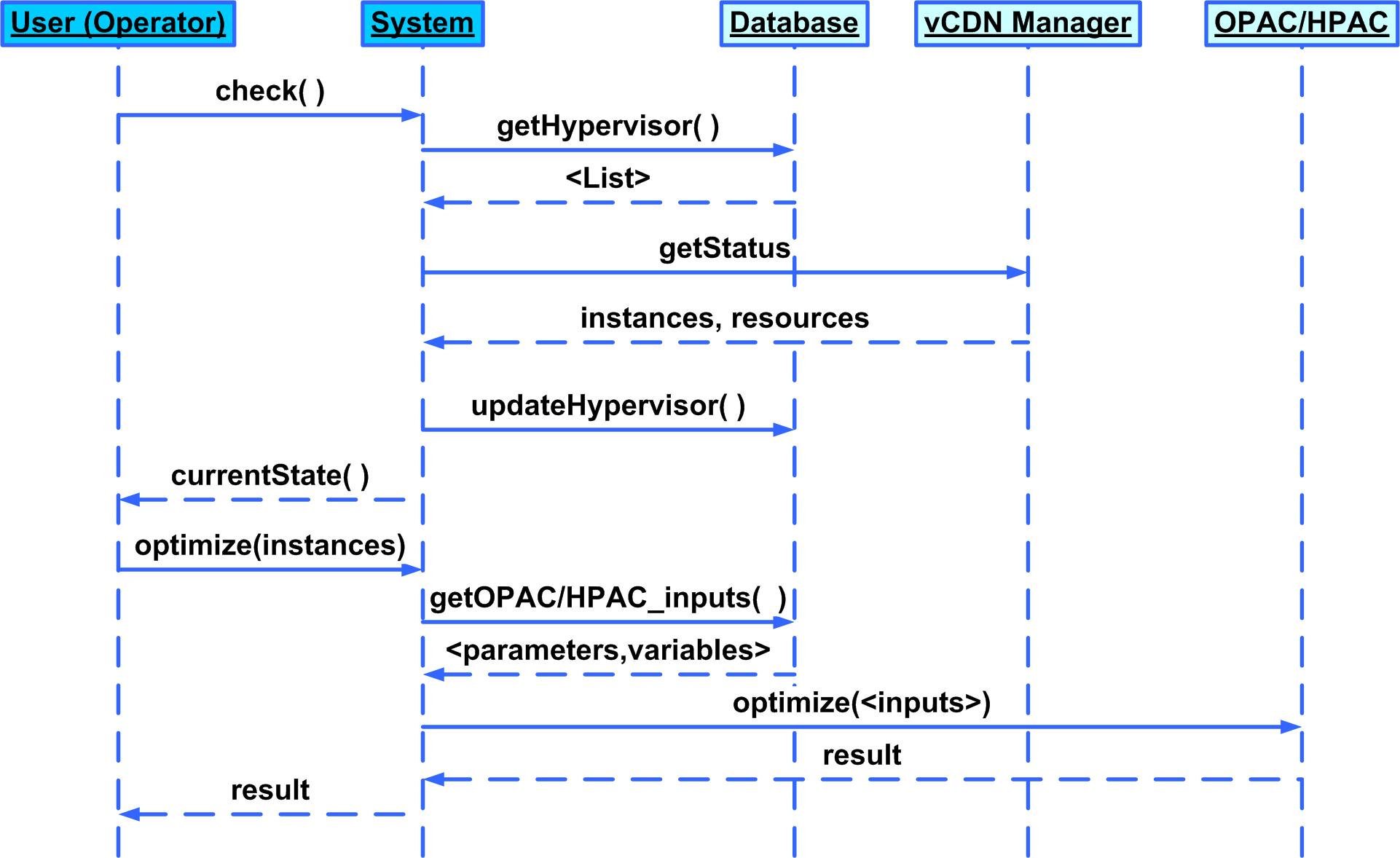}
\caption{Sequence diagram for OPAC/HPAC integration.}
\label{sesuence-diagram}
\end{center}
\end{figure}

From an operator perspective, integrating the proposed algorithms may follow the sequence diagram represented by the Fig. \ref{sesuence-diagram}. Indeed, the main stakeholders are:
\begin{itemize}
\item User (Operator): Operator of the infrastructure, deciding the migration or not of a vCDN instance according to the current state of the environment and the result given by the optimization algorithms OPAC/HPAC.
\item System: This system, offering the Users an interaction with the OPAC/HPAC algorithms inputs and results. It still maintains the operator database updated by the necessary information about the  operator's servers (\textit{getHypervisor()}) and the provider's vCDNs (\textit{getStatus()}).
\item Database: It contains the information about the SDN/NFV infrastructure status and values. In this case, it is an SQLite database.
\item vCDN Manager: It takes care of the interaction with the software managers running in the architecture, to poll from them the required information for the OPAC/HPAC algorithm. It will use internally API calls to the different software managers (Openstack Horizon, Opendaylight, etc.).
\item OPAC and HPAC Algorithms:
Implementation of the optimization algorithms that optimize the vCDN placement and migration based on exact/heuristic optimization approaches. A file is given as an input and a file is returned as an output of the selected algorithm. The algorithm selection depends mainly on the network scale. 
\end{itemize}

For simplicity, when the user (operator) executes the optimize command, the system fetches the operator databases to get the required information \textit{$(getOPAC/HPAC\_input())$}in order to launch in turn the optimize command. Then, OPAC/HPAC algorithm decides where to place and migrate the vCDNs and provide the system/operator the result. Finally, the system executes the migration process according to our results and releases the dedicated resources in case of Service Level Agreement (SLA)-expiration/vCDN-delete-request.

Recall that the proposed algorithms involved also the content provider who wanted to rent a cloud of vCDN for its customers. Therefor, from a content provider perspective, the main exchanges between the content provider, the vCDN manager and the user (operator) are depicted:
\begin{itemize}
\item The content provider (e.g., YouTube) requests a vCDN creation during a specific time (e.g., 2 hours) and with a specific QoE (e.g., excellent) and to cover a specific region (e.g. Paris).
\item The network operator, representing the owner of the NFV infrastructure, checks the current state of its NFV resources and call the proposed optimization algorithms through the provided \textit{system}.
\item This system interacts with a operator database server to retrieve the required information about network topology, NFV resources and update another database dedicated to the placement/migration decision.
\item The NFV Infrastructure (NFVI) administrator interacts with the decision's database to migrate the resources while keeping the signed SLA between the content provider and the network operator valid and standing.
\item Optionally, the NFVI administrator may let the content provider as the manager of the video content.
\item The decision database is updated by the operator for security issues.
\item In case of SLA expiration (e.g, covering time expiration, vCDN delete-request, etc..), the operator releases the dedicated NFV resources.
\end{itemize}

\section{Conclusion}
This paper presents two optimization solutions either for the placement or the migration problem of virtual CDNs. Multiple constraints that are strongly related to the CDN virtualization are taken into account such as vCDN size, content resolution and system/network requirements. In addition, the two optimization algorithms OPAC and HPAC target respectively  different network scales. Then, they are modeled and implemented. In small scale networks, results show that $\left\vert{F}\right\vert$ have more significant impact on the average migration cost in the case of using HPAC rather than OPAC. Nevertheless, it is noticeable that HPAC converges to the optimal solution and outperforms OPAC in specific metrics. In large scale, HPAC gives a short execution time (e.g. the run-time was in terms of a few seconds) which proofs its efficiency and scalability. In this network scale, The HPAC's results show that $\left\vert{F}\right\vert$ have insignificant impact on the migration time, the migration cost and the replication number since the system reaches quickly its stability. Moreover, the two approaches are integrated in an SDN/NFV framework and the main use case, sequence diagram from either an operator or content provider perspective. Future work can be conducted on integrating the aforementioned placement algorithms using Information-Centric Networking (ICN) context regarding its ubiquitous and in-networking caching features. 
\ifCLASSOPTIONcaptionsoff
  \newpage
\fi
 
 \IEEEtriggeratref{8}
 \IEEEtriggercmd{\enlargethispage{-5in}}
\bibliographystyle{IEEEtranS}
\bibliography{IEEEabrv,references}

\end{document}